\documentclass{article}

\newif\ifIEEEcompliantpdf
\IEEEcompliantpdffalse

\makeatletter

\usepackage{spconf}
\usepackage{bm,amssymb,amsmath,mathrsfs}
\usepackage{cite}
\usepackage{algorithm,algorithmic}
\usepackage{tabularx,colortbl,makecell}
\usepackage[tableposition=top]{caption}
\usepackage[hang,flushmargin]{footmisc}

\usepackage{paralist}
\usepackage{enumitem}
\setlist{noitemsep} 

\usepackage[T2A,T1]{fontenc}
\usepackage{txfonts}
\usepackage{mathptmx}

\usepackage[utf8]{inputenc}
\usepackage[russian,english]{babel}
\usepackage{substitutefont}

\DeclareRobustCommand{\cyrins}[1]{%
  \begingroup\hyphenpenalty\@M\exhyphenpenalty\@M\fontfamily{Tempora-TLF}%
  \foreignlanguage{russian}{#1}%
  \endgroup }

\usepackage{color}
\usepackage{graphicx}

\ifIEEEcompliantpdf
\def\href#1#2{#2}%
\else
\RequirePackage[pdftex,pdfpagemode=none, pdftoolbar=true,
                pdffitwindow=true,pdfcenterwindow=true]{hyperref}
\fi

\definecolor{darkgreen}{rgb}{0,0.5,0}
\definecolor{darkyellow}{rgb}{0.5,0.5,0}
\definecolor{darkred}{rgb}{0.6667,0,0}
\definecolor{light-gray}{gray}{0.5}

\def\twoauthors#1#2#3#4{\gdef\@address{}
   \gdef\@name{\begingroup\normalsize\begin{tabular}{@{}c@{}}
        {\large\em #1}\\\noalign{\vskip 6pt plus 3pt minus 3pt}
        #2\relax
   \end{tabular}\hskip 0.3333in plus.1667in minus.1667in\begin{tabular}{@{}c@{}}
        {\large\em #3}\\\noalign{\vskip 6pt plus 3pt minus 3pt}
        #4\relax
\end{tabular}\endgroup}}

\def\threeauthors#1#2#3#4#5#6{\gdef\@address{}
   \gdef\@name{\begin{tabular}{@{}c@{}}
        {\em #1}\\\noalign{\vskip 6pt plus 3pt minus 3pt}
        #2\relax
   \end{tabular}\hskip .5in plus.5in minus.125in\begin{tabular}{@{}c@{}}
        {\em #3}\\\noalign{\vskip 9pt plus 3pt minus 3pt}
        #4\relax
   \end{tabular}\hskip .5in plus.5in minus.125in\begin{tabular}{@{}c@{}}
        {\em #5}\\\noalign{\vskip 6pt plus 3pt minus 3pt}
        #6\relax
\end{tabular}}}

\def\thinnerspace{\kern .1111em }

\def\betab{{\bm{\beta}}}
\def\gammab{{\bm{\gamma}}}

\def\Ab{{\mathbf{A}}}

\def\Fb{{\mathbf{F}}}

\def\xbmit{{\bm{x}}}
\def\ybmit{{\bm{y}}}

\DeclareMathOperator{\Normal}{Normal}

\DeclareMathOperator{\Expect}{\mit E}

\mathchardef\colonord="003A

\def\T{{\scriptscriptstyle\rm T}}   
\let\humlaut=\H
\def\H{\ifmmode{\scriptscriptstyle\rm H}\else\humlaut\fi}   

\def\by{\ifmmode $\hbox{-by-}$\else \leavevmode\hbox{-by-}\fi}
\def\sqrtm1{{\sqrt{\!-1}}}
\let\(=\langle
\let\)=\rangle

\def\hgfarg#1{\left(\null\vcenter{\normalbaselines\m@th
    \ialign{&$\displaystyle##$\hfil\crcr
      \mathstrut\crcr\noalign{\kern-\baselineskip}
      #1\crcr\mathstrut\crcr\noalign{\kern-\baselineskip}}}\right)}
\def\Heqalign#1{\null\,\vcenter{\openup\jot\m@th
  \ialign{\strut\hfil$##$:\quad&\hfil$\displaystyle{##}$&$\displaystyle
      {{}##}$\hfil&\qquad##\hfil\crcr#1\crcr}}\,}
\def\eqalignno#1{\displ@y \tabskip\@centering
  \halign to\displaywidth{\hfil$\@lign\displaystyle{##}$\tabskip\z@skip
    &$\@lign\displaystyle{{}##}$\hfil\tabskip\@centering
    &\llap{$\@lign##$}\tabskip\z@skip\crcr
    #1\crcr}}
\def\dbleqalignno#1{\displ@y \tabskip\@centering
  \halign to\displaywidth{\hfil$\@lign\displaystyle{##}$\tabskip\z@skip
    &$\@lign\displaystyle{{}##}$\hfil
    &$\@lign\displaystyle{{}##}$\hfil\tabskip\@centering
    &\llap{$\@lign##$}\tabskip\z@skip\crcr
    #1\crcr}}

\newif\ifmathtomb \mathtombfalse

\def\tombstone{\unskip\penalty50   
  \hskip 0pt plus-1fill \null\nobreak\hskip 0pt plus1fill
  \enskip \vrule width.3333em height.7em depth.2em
  \ifmmode \global\mathtombtrue \else \global\mathtombfalse \fi}

  {\futurelet\next\hpr@oftext}
  {\ifmathtomb \else \tombstone \fi \widowpenalty=10000  
   \par \ifmathtomb \else \addvspace{\medskipamount}\fi \global\mathtombfalse}
\def\hpr@oftext{\ifx\next[\let\temp\ohpr@@ftext\else\let\temp\hpr@@ftext\fi\temp}
\def\hpr@@ftext{\beginhpr@@f{Proof}}
\def\ohpr@@ftext[#1]{\beginhpr@@f{#1}}
\def\beginhpr@@f#1{\par \addvspace{\bigskipamount}%
  \noindent{\bf #1:\enspace}\ignorespaces }

\def\twittersize{\small}
\def\twitter#1{{\twittersize\textsf{#1}}}
\definecolor{twitterblue}{rgb}{0.9025, 0.9363, 0.9625}

\usepackage{etoolbox}
\newlength{\qrr@dimen@}
\expandafter\pretocmd\csname tabular*\endcsname{\setlength{\qrr@dimen@}{#1}}{}{}
\newcommand*{\Rowcolor}[2][\tabcolsep]{%
    \ifx\relax#1\relax\else
        \kern-\the\dimexpr#1\relax
    \fi
    \makebox[0pt][l]{%
        \fboxsep=0pt
        \colorbox{#2}{%
            \strut\kern\qrr@dimen@
        }%
    }%
    \ifx\relax#1\relax\else
        \kern\the\dimexpr#1\relax
    \fi
    \ignorespaces
}

\newsavebox\saved@arstrutbox
\newcommand*{\setarstrut}[1]{%
  \noalign{%
    \begingroup
      \global\setbox\saved@arstrutbox\copy\@arstrutbox
      #1%
      \global\setbox\@arstrutbox\hbox{%
        \vrule \@height\arraystretch\ht\strutbox
               \@depth\arraystretch \dp\strutbox
               \@width\z@
      }%
    \endgroup
  }%
}
\newcommand*{\restorearstrut}{%
  \noalign{%
    \global\setbox\@arstrutbox\copy\saved@arstrutbox
  }%
}

\makeatother

\title{Influence Estimation on Social Media Networks Using Causal Inference}
\twoauthors{Steven T.~Smith, Edward K.~Kao, Danelle C.~Shah, Olga Simek,*
and}
{MIT Lincoln Laboratory\\
  Lexington MA 02421 USA\\ \{\thinspace\href{mailto:stsmith@ll.mit.edu}{stsmith},
  \href{mailto:edward.kao@ll.mit.edu}{\hbox{edward.kao}},
  \href{mailto:danelle.shah@ll.mit.edu}{danelle.shah},
  \href{mailto:osimek@ll.mit.edu}{osimek}\thinspace\}@ll.mit.edu}
{Donald B.~Rubin*}{Harvard University\\
  Cambridge MA 02138 USA\\ \href{mailto:rubin@stat.harvard.edu}{rubin@stat.harvard.edu}}

\begin{document}


\topmargin 0truept
\headheight 0truept
\headsep 0truept
\textheight 229truemm

\textwidth 507truept

\twocolumn
\columnsep 6truemm

\def\ninept{\def\baselinestretch{.95}%
  \let\Huge\huge
  \let\huge\LARGE
  \let\LARGE\large
  \let\Large\large
  \let\large\normalsize
  \let\normalsize\small
  \let\small\footnotesize
  \let\footnotesize\scriptsize
  \let\scriptsize\tiny
  \normalsize
}
\ninept

\maketitle

\begin{abstract} Estimating influence on social media networks is an
important practical and theoretical problem, especially because this
new medium is widely exploited as a platform for disinformation and
propaganda. This paper introduces a novel approach to influence
estimation on social media networks and applies it to the real-world
problem of characterizing active influence operations on Twitter
during the 2017 French presidential elections. The new influence
estimation approach attributes impact by accounting for narrative
propagation over the network using a network causal inference
framework applied to data arising from graph sampling and
filtering. This causal framework infers the difference in outcome as a
function of exposure, in contrast to existing approaches that
attribute impact to activity volume or topological features, which do
not explicitly measure nor necessarily indicate actual network
influence.  Cram\'er-Rao estimation bounds are derived for parameter
estimation as a step in the causal analysis, and used to achieve
geometrical insight on the causal inference problem. The ability to
infer high causal influence is demonstrated on real-world social media
accounts that are later independently confirmed to be either directly
affiliated or correlated with foreign influence operations using
evidence supplied by the U.S.~Congress and journalistic reports.
\let\thefootnote\relax\footnotetext{\noindent *This material is based
  upon work supported by the Assistant Secretary of Defense for
  Research and Engineering under Air Force Contract
  No.\ FA8721-05-C-0002 and\slash or FA8702-15-D-0001. Any opinions,
  findings, conclusions or recommendations expressed in this material
  are those of the authors and do not necessarily reflect the views of
  the Assistant Secretary of Defense for Research and Engineering.}
\end{abstract}

\keywords Estimation on graphs, influence estimation, causal
inference, social networks, graph analytics, network detection,
Cram\'er-Rao bounds on graph estimation\endkeywords

\section{INTRODUCTION}
\label{sec:intro}

Social media is now used by a majority of individuals across the
industrialized world, and in 2018 is estimated to overtake traditional
media in the U.S.~\cite{Statista-nd}. This large-scale growth of
worldwide social media networks is built upon their essential features
of universal access, immediacy, and power to communicate with and
influence others. These key design features have also created a potent
new medium and an enabling technology for disinformation and
propaganda~\cite{Confessore2018,Fan2014,Shah2011,Stewart2018,Tambuscio2015,Zhang2015}.
Detecting and estimating influence on social media networks is the
problem of inferring the impact of an input at one node on the rest of
the network. This is an important theoretical and practical problem
that arises in marketing on social media \cite{Confessore2018,Li2014},
influence maximization \cite{Chen2009,Kempe2015}, information
diffusion \cite{Kimura2006,Leskovec2007,Myers2012,Zhang2015}, and the
spread of both information and disinformation in social networks
\cite{Budak2011,Confessore2018,Fan2014,Jin2013,Nguyen2012,Shah2011,Stewart2018,Tambuscio2015,Wen2015}.

This paper introduces a novel approach to influence estimation on
social media networks and applies it to the real-world problem of
characterizing online influence
operations~\cite{Borger2017,Chekinov2013,Rogers2017}.  The new
framework contains the generality to account for influence on
populations with and without observed outcomes, which addresses the
issue of biased sampling. Both groups contain individuals that are
receptive to influence, or not receptive, the latter defined as having
outcomes invariant to the influence exposure. The approach also
subsumes the important case of a network engaged in a specific
narrative.  The estimation problem is posed using Bayesian inference
with these distinctive aspects of social media networks:
\begin{itemize}
\item (Sampled) graph data of social network interactions
\item Heterogeneous natural language and multi-media data
\item Causal relationship among interactions
\end{itemize}

\subsection{Contributions}
\label{sec:contrib}

The new influence estimation approach proposed here aims to attribute
impact by accounting for causal narrative propagation over the
network, while addressing the challenge of discriminating between
actual social influence and mere
homophily~\cite{Shalizi2011,Stewart2018}.  Statistical inferential
methods are used to estimate model parameters for causal influence on
social networks.  This causal framework infers the difference in
outcome as a function of exposure, in contrast to existing approaches
that attribute impact to activity volume or topological features,
which do not explicitly measure nor necessarily indicate actual
network influence.  Cram\'er-Rao estimation bounds are derived for
this parameter estimation problem and used for geometrical insight on
the causal inference problem.  Natural language processing tools are
used to filter the network data into specific narrative contexts for
influence estimation, and to classify the sampled graph data into
case\slash non-case data.

This approach is applied to publicly available Twitter data collected
over the 2017 French presidential elections, during which there was an
active influence operation campaign targeting these
elections~\cite{Borger2017,Chekinov2013,Rogers2017}.  The ability to
infer high causal influence is demonstrated on real-world social media
accounts that were later independently confirmed to be either directly
affiliated or correlated with foreign influence operations by evidence
supplied by the U.S.~Congress~\cite{USHPSCI2017} and journalistic
reports~\cite{Marantz2017}. Furthermore, the new approach is shown to
reveal influential accounts that are not obvious based upon simple
activity statistics. These results support the utility of the proposed
framework.

Section~\ref{sec:graph-sampling} describes the network sampling
procedure.  Section~\ref{sec:graph-filtering} describes the narrative
context filtering approach. The novel approach to causal influence
estimation is given in Section~\ref{sec:graph-influence}. Real-world
data examples will be used to illustrate the overall approach
throughout the paper.

\section{GRAPH SAMPLING USING TARGETED CONTEXT}
\label{sec:graph-sampling}

The vast majority of social media interactions are irrelevant to a
intended context of interest for influence estimation.  Therefore, an
appropriate graph sampling mechanism is needed.  Ideally, the sampling
mechanism must efficiently capture relevant examples of both the
influence network and associated content, under rate-limited server
queries.  In this paper we use a content-based approach for network
graph sampling based on a method for optimum network
detection~\cite{Smith2014,Smith2017}.

\begin{figure}[t]
\normalsize
\centerline{\includegraphics[width=1.0\linewidth]{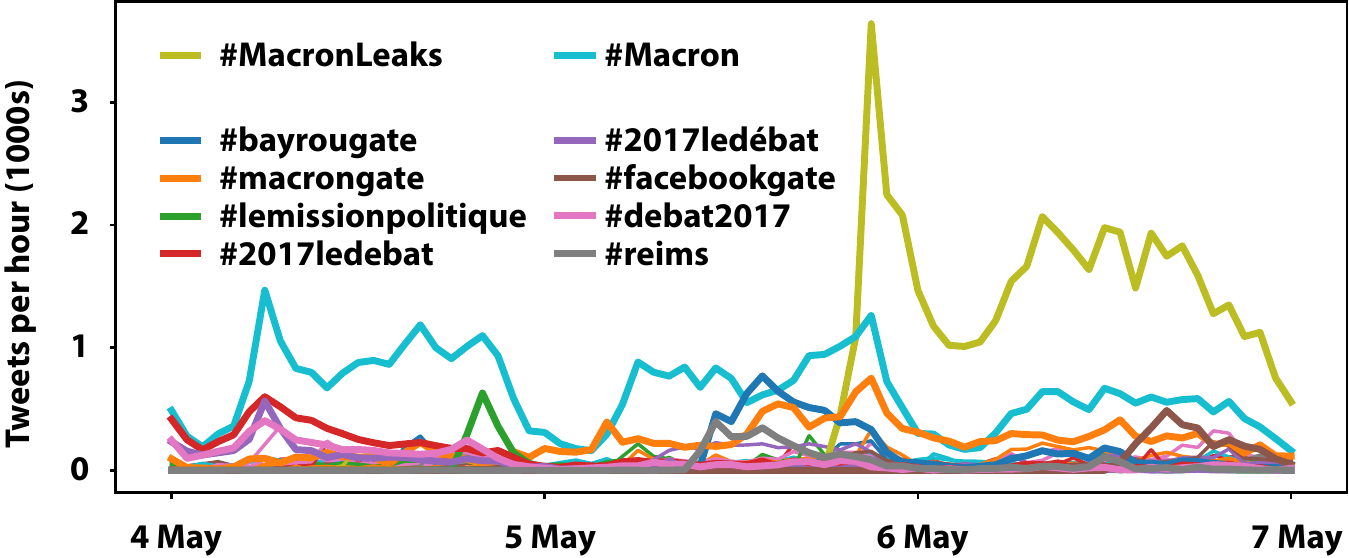}}
\caption{Hashtag frequencies during the 2017 French elections.  The
  \twitter{\#MacronLeaks} narrative, referring to thousands of leaked
  documents, dominated election-related social media discussions hours
  before the mandatory French media
  blackout~\cite{Borger2017,Marantz2017}.\label{fig:2017frenchhastags}}
\end{figure}

First, potentially relevant social media content (e.g.\ hashtags or
keywords) and user account handles are identified using prior
knowledge and subject matter expertise about the targeted context,
similar to Berger's approach~\cite{Berger2017}. Second, publicly
available information associated with these prior cues are collected.
Relevance is determined by a set of hashtags, keywords, and user
account handles whose content is aligned with hypothesized influence
campaigns. Third, graph sampling is performed by selectively
collecting content prioritized by a threat propagation model on the
graph~\cite{Smith2014,Smith2017}.

\begin{figure}[t]
\normalsize
\centerline{\includegraphics[width=1.0\linewidth]{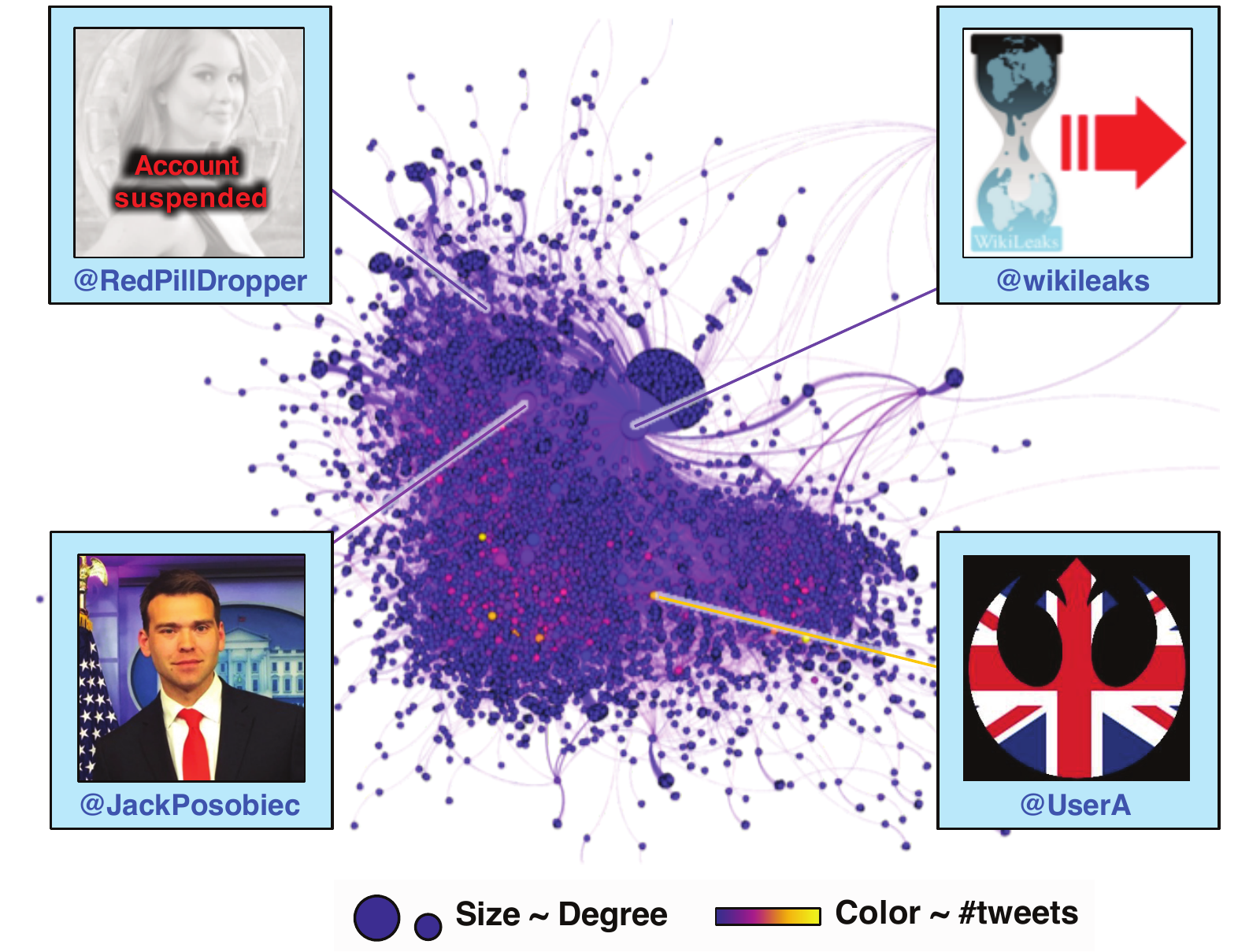}}
\caption{Social network of the \twitter{\#MacronLeaks} narrative,
  5--7~May, highlighting potentially influential accounts based on
  journalistic reports and high
  activity~\cite{Marantz2017}.\label{fig:macronleaksgraph}}
\end{figure}

More than 20~million tweets and 2~million user account handles of
potential relevance to foreign influence operations in the 2017 French
presidential elections were collected using the Twitter public
application programming interface (API)~\cite{Twitter2018} during
April and May 2017.  This data set includes accounts that are
positively identified as part of the foreign influence
campaign~\cite{USHPSCI2017}; however, no affiliation is implied or
ascribed to any specific account without independent
corroboration. Fig.~\ref{fig:2017frenchhastags} shows a plot of the
top Twitter hashtag narrative frequencies collected during this
period. The prominent \twitter{\#MacronLeaks} narrative, referring to
thousands of leaked documents, dominated election-related social media
discussions hours before the mandatory French media blackout for the
second round on 7~May~\cite{Borger2017,Marantz2017}. In
Section~\ref{sec:graph-influence}, the network associated with this
narrative will be used to demonstrate estimation of the most
influential accounts.  Fig.~\ref{fig:macronleaksgraph} illustrates
this narrative graph (5,370 accounts, 124,259 interactions) and
highlights potentially influential accounts based upon activity levels
and subsequent journalistic reports~\cite{Marantz2017}. Some of these
accounts will be shown to possess high levels of causal influence,
whereas others are estimated to have much less influence, underscoring
the insight that high activity is not necessarily correlated with high
influence.

\section{GRAPH FILTERING USING NARRATIVES}
\label{sec:graph-filtering}

\begin{figure}[t]
\normalsize
\centerline{\includegraphics[width=0.8\linewidth]{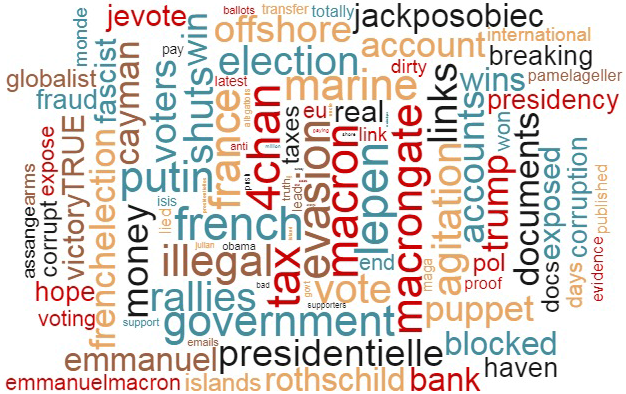}}
\caption{Tag cloud topic narrative of the \twitter{\#MacronLeaks}
  hashtag.\label{fig:macronleakscloud}}
\end{figure}

Social influence occurs within a specific context. The sampled data
must be filtered by a characterization of the narrative prior to
influence estimation. This may be as simple as searching for a single
keyword, utilizing natural language processing to focus on a specific
narrative, or community detection approaches to filter the graph, or
some combination of these. E.g.\ Stewart et~al.\ use an Infomap
clustering algorithm to perform community detection on a retweet
graph~\cite{Stewart2018}. One narrative derived from the content of
the \twitter{\#MacronLeaks} tweets~\cite{Borger2017} is illustrated by
the wordcloud in Fig.~\ref{fig:macronleakscloud}. For example, an
account \twitter{@Pamela\_Moore13} known to be tied to foreign
influence operations~\cite{USHPSCI2017} is observed to promote this
and many other hashtag-based divisive narratives aligned with these
operations. \begingroup\predisplaypenalty=0
$$\includegraphics[width=1.0\linewidth]{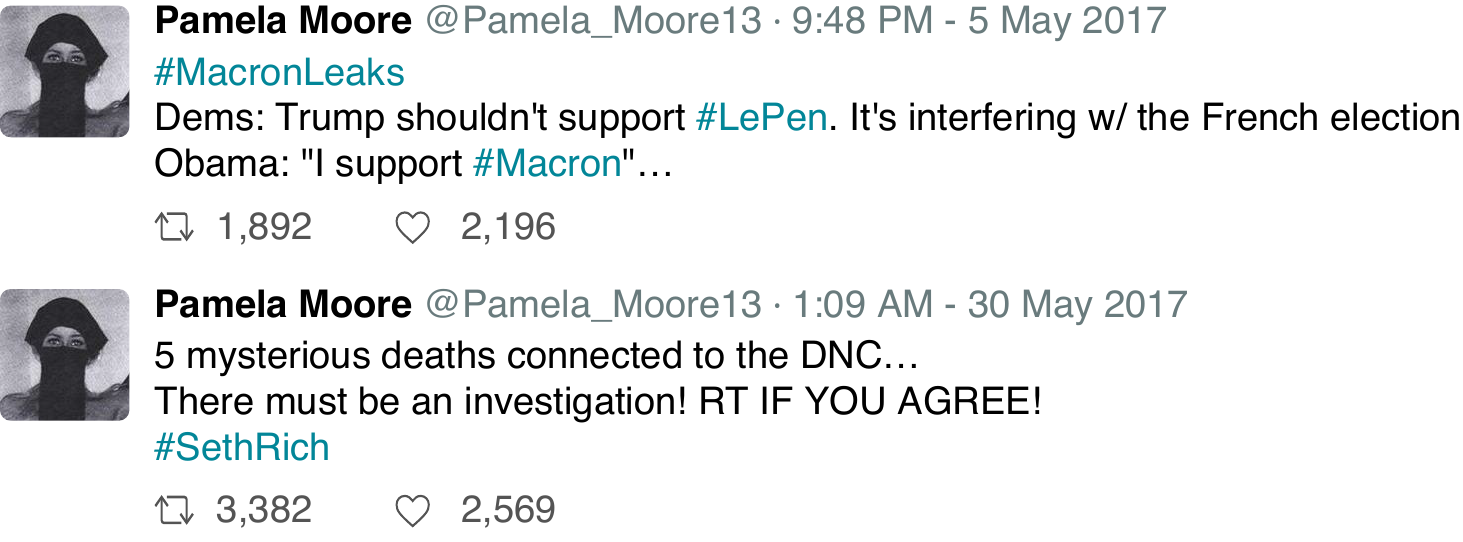}$$
The \twitter{\#MacronLeaks} narrative example is prominent and
well-defined; therefore, a simple keyword search is an adequate
filter. In general, combining content-based narrative detection with
community detection is expected to be necessary to estimate influence
in broader contexts than provided by a single hashtag. \par\endgroup

\section{GRAPH INFLUENCE ESTIMATION}
\label{sec:graph-influence}

\begin{figure}[t]
\normalsize
\centerline{\includegraphics[width=1.0\linewidth]{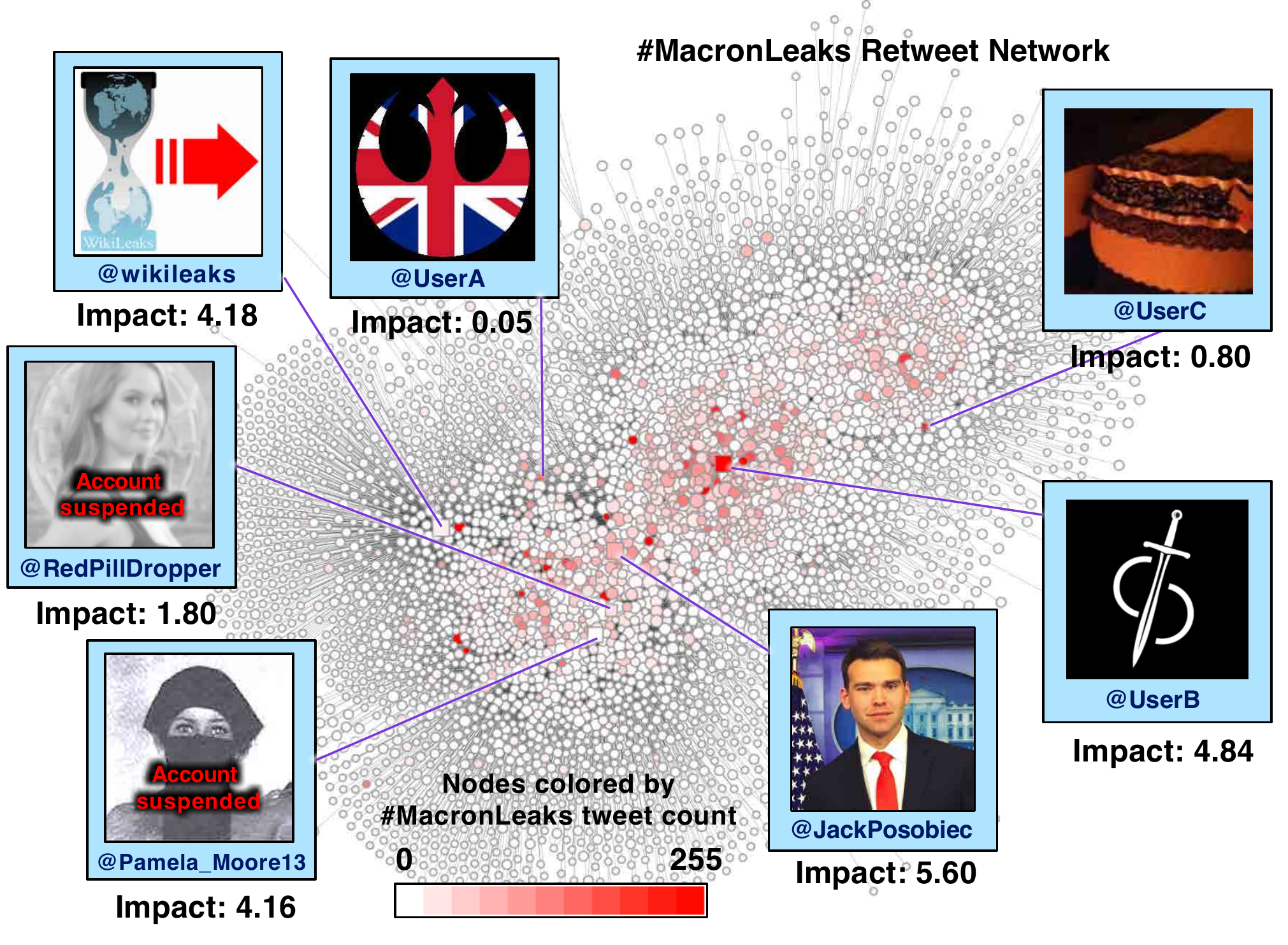}}
\caption{Influence estimation on the \twitter{\#MacronLeaks} narrative
  network of Fig.~\ref{fig:macronleaksgraph}. Vertices are accounts
  and edges are retweets. Vertex color indicates the number of tweets
  and the vertex size corresponds to out-degrees. Impact is the
  average number of additional tweets generated by an user's
  participation
  [Eq.~(\ref{eq:individual_impact_causal_estimand})].\label{fig:influencemacronleaks}}
\end{figure}

The proposed causal framework for influence estimation introduced in
this section infers the difference in outcome as a function of
exposure, in contrast to approaches that attribute impact to activity
volume or topological features.  Existing approaches are based on
network topology~\cite{Kimura2006,Leskovec2007,Shah2011}, node degree
or activity~\cite{Nguyen2012,Wen2015}, information
diffusion~\cite{Kempe2015,Li2014,Myers2012}, and cascade
lengths~\cite{Budak2011,Chen2009,Fan2014,Leskovec2007,Li2014}.  None
of these metrics explicitly measure nor necessarily indicate actual
network influence.  For example, an account that tweets little can be
highly influential, and vice-versa. Examples of this claim are
demonstrated with real data below.  The proposed causal inference
approach aims to quantify influence by accounting for causal narrative
propagation over the entire network, including the timing of the
tweets and the position of the influencer in the network. It also
accounts for several potential confounders (e.g.\ community
membership, popularity) and removes their effects from the causal
estimation.  The impact of an individual account on the narrative
network [Fig.~\ref{fig:macronleaksgraph}] is defined as the average
(per~vertex) number of additional tweets generated by its
participation [Fig.~\ref{fig:influencemacronleaks}]. This impact is
estimated using a recent network causal inference
framework~\cite{Kao2017}, itself based upon Rubin's causal framework
\cite{ImbensRubin2015}. This approach is related to other work in
causal inference on
networks~\cite{Sussman2017,Toulis2013,Ugander2013}; it is unique in
its inference methodology that uses the network potential outcomes.

\subsection{Causal Influence Estimation on Networks}
\label{sec:causal-influence-problem}

Let $G=(V,E)$ be a graph with $N$ vertices $V=\{v_1,v_2,\ldots,v_N\}$,
whose edges are denoted by the observed interactions between $v_i$
and~$v_j$, let $\Ab=(a_{ij})$ be an $N\by N$ matrix of social
influence of~$v_i$ on~$v_j$ with Poisson rate determined by the graph
data~$G$, and $\bm{Z}$ be a binary $N$-vector of narrative sources
(a.k.a.\ treatment vector). The fundamental quantity is the network
potential outcome of each individual vertex, denoted
$Y_i(\bm{Z},\Ab)$, under exposure to the narrative through the source
vector $\bm{Z}$ and influence network $\Ab$.  In the analysis below,
vertices are user accounts in the narrative network, edges are
retweets relevant to the narrative, and the potential outcomes are the
number of tweets in the narrative. The impact $\zeta_i$ of each vertex
on the narrative is defined using the network potential outcomes,
\begin{equation}
\label{eq:individual_impact_causal_estimand}
	\zeta_i(\bm{z}) \buildrel{\rm def}\over= \frac{1}{N} \sum_{j =
          1}^N \bigl(Y_j(\bm{Z}=\bm{z}_{i+}, \Ab)
        -Y_j(\bm{Z}=\bm{z}_{i-}, \Ab)\bigr).
\end{equation}
This causal estimand is the average difference between the individual
potential outcomes with $v_i$ as a source
s.t.\ $\bm{z}_{i+}=(z_1,\ldots,{z_i:=1},\discretionary{}{}{}\ldots,z_N)^\T$,
versus $v_i$ {\em not\/} a source
s.t.\ $\bm{z}_{i-}=(z_1,\ldots,{z_i:=0},\discretionary{}{}{}\ldots,z_N)^\T$.
This impact is the average (per~vertex) number of additional tweets
generated by a vertex $v_i$'s participation. The source is {\em
  uniquely impactful\/} if it is the only source.

It is impossible to observe the potential outcomes at each vertex with
both exposure conditions under source vectors $\bm{z}_{i+}$ and
$\bm{z}_{i-}$; therefore, the missing potential outcomes must be
estimated. This can be accomplished by modeling the potential
outcomes. After estimating the model parameters from the observed
outcomes and vertex covariates, missing potential outcomes in the
causal estimand $\zeta_i$ can be imputed using the fitted model.

In this analysis, potential outcomes are modeled using a conditional
Poisson generalized linear model (GLM) with the canonical log link
function and linear predictor coefficients $(\tau,\gammab,\betab,\mu)$
corresponding to the source indicator $Z_i$, $n$-hop exposure vector
$s^{(n)}_i(\bm{Z})$, the covariate vector $\xbmit_i$, and the baseline
outcome. The first two parameters $\tau$ and~$\gammab$ account for
influence from the source(s), and the second two parameters $\betab$
and~$\mu$ account for individual baseline propensity.  The covariate
vector $\xbmit_i$ has $m$ elements corresponding to the number of
observed or inferred covariates on each vertex, including the
potential social confounders such as popularity and community
membership. The conditional GLM model for the potential outcomes is,
\begin{equation}
\label{eq:PO_PoissonGLM}
	\log\lambda_i = \tau Z_i + \biggl(\sum_{n=1}^{N_{\text{hop}}}
                   { \prod_{k=1}^n{\tau \gamma_k s^{(n)}_i(\bm{Z})}}\biggr) +
                   \betab^\mathrm{T} \bm{x}_i + \mu + \epsilon_i,
\end{equation}
with $Y_i(\bm{Z},\Ab) \sim \text{Poisson}(\lambda_i)$. The first term
in the linear predictor $\tau Z_i$ represents the primary effect of
the source. The second term represents the accumulative social
influence effect from $n$-hop exposures $s^{(n)}_i(\bm{Z})$ to the
source, where each coefficient $\gamma_k$ captures the decay of the
effect over each additional hop. The third term in
Eq.~(\ref{eq:PO_PoissonGLM}) is the effect of the vertex covariates
$\bm{x}_i$ including the potential social confounders such as
popularity and community membership. In the analysis below, two
confounders are used: popularity, based on vertex degree, and
community membership, based on language.  The fourth term, $\mu$, is
the baseline effect on the entire population. The last term
$\epsilon_i \sim \Normal(0,\sigma_\epsilon^2)$ gives independent and
identically distributed variation for heterogeneity between the
units. The amounts of social exposure at the $n$th hop are determined
by $(\Ab^\T)^n\bm{Z}$. This captures all exposure to the sources
propagated on the influence network. Lastly, to model the diminishing
return of additional exposures, the (nonnegative) log-exposure is
used, $\bm{s}^{(n)}(\bm{Z}) = \log\bigl((\Ab^\T)^n \bm{Z} + 1\bigr)$.

Joint Bayesian inference of the model parameters $\tau$, $\gammab$,
$\betab$, $\mu$, and the social influence matrix $\Ab$ is done through
Markov~chain Monte~Carlo (MCMC) with Bayesian regression
updates~\cite{Gelman2013}, in which the observations
$\bm{Y}(\bm{Z}=\bm{z}_{i+}, \Ab)$ are used to compute samples of the
posterior density $P(\tau,\gammab,\betab,\mu\mid\bm{Y})$. The samples
are used to impute potential outcomes via
Eq.~(\ref{eq:PO_PoissonGLM}). Gibbs sampling is used to sequentially
update individual model parameter estimates, conditioned on estimates
of the others.  Weakly informative, truncated priors are chosen to
reduce the prior assumption's effect on the posterior density, and to
improve MCMC robustness and convergence.

\subsection{Influence Estimation Accuracy}
\label{sec:influence-accuracy}

The lower bound on the sampling variance of the influence estimate is
an important quantity to compute because it indicates both the best
precision achievable, and provides geometric insight into the
dependency of the influence estimate on both the data and
parameters~\cite{Smith2005}. The Cram\'er-Rao lower bound on the model
parameter sampling covariance is computed from the inverse of the
Fisher information matrix of the proposed Poisson
GLM, \begin{equation}\Fb = -\Expect_{\ybmit}\bigl[
    \partial^2\ell(\tau, \gammab, \betab, \mu)/\partial(\tau, \gammab,
    \betab, \mu)^2\bigr],
\end{equation}
where $\ybmit$ are the observed potential outcomes and $\ell$ is the
log-likelihood function.  For the simplest form of the proposed
Poisson-GLM model with $1$-hop exposure, a scalar vertex covariate,
and conditional independence between vertices, the Fisher information
matrix is,
\begin{equation}
\label{eq:FisherInformationMatrix}
\Fb = \sum_{i=1}^N\lambda_i\begin{pmatrix} \phi_i^2&
\phi_i\tau s_i^{(1)}& \phi_ix_i&
\phi_i\\ &(\tau s_i^{(1)})^2 &\tau
s_i^{(1)} x_i &\tau s_i^{(1)}\\ &&x_i^2&x_i\\ &&&1\end{pmatrix},
\end{equation}
where $\phi_i\buildrel{\rm def}\over=Z_i+\gamma_1 s_i^{(1)}$.

\begin{table}[t]
\setlength{\tabcolsep}{0.15em}
\newcommand\TopStrut{\rule{0pt}{2.6ex}}       
\newcommand\BotStrut{\rule[-1.2ex]{0pt}{0pt}} 
\newcolumntype{S}{>{\small}r}
\newcolumntype{T}{>{\small\let\twittersize\footnotesize}l}
\caption{\twitter{\#MacronLeaks} narrative network screen names,
  tweets (T), total retweets (TRT), most retweeted tweet (MRT),
  followers (F) [in thousands], initial times (on 5~May), PageRank
  centrality (PR)~\cite{Langville2005,Peixoto2014}, and estimated
  influence of example accounts.}
\label{tab:macronleaks_screennames}
\begin{tabular*}{\linewidth}{T@{\extracolsep{\fill}}SSSSSSS}
\hline
\multicolumn{1}{l}{Screen name} & \multicolumn{1}{c}{\ T} &
  \multicolumn{1}{c}{TRT} & \multicolumn{1}{c}{MRT} &
  \multicolumn{1}{c}{\quad F} &
  \multicolumn{1}{c}{1st time} &
  \multicolumn{1}{c}{\quad PR} &
  \multicolumn{1}{r}{\bf Impact*\TopStrut\BotStrut}\\
\hline
\Rowcolor{twitterblue}\twitter{@JackPosobiec} & 95 & 47\thinnerspace k & 5\thinnerspace k & 261\thinnerspace k & 18:49 & 2.84 &\bf 5.60\TopStrut \\
\twitter{@RedPillDropper} & 32 & 8\thinnerspace k & 3\thinnerspace k & 8\thinnerspace k & 19:33 & 2.86 &\bf 1.80 \\
\Rowcolor{twitterblue}\twitter{@UserA}\dag & 256 & 59\thinnerspace k & 8\thinnerspace k & 1\thinnerspace k & 19:34 & 27.08 &\bf 0.05 \\
\twitter{@UserB}\dag & 260 & 54\thinnerspace k & 8\thinnerspace k & 3\thinnerspace k & 20:25 & 57.05 &\bf 4.84 \\
\Rowcolor{twitterblue}\twitter{@wikileaks} & 25 & 63\thinnerspace k & 7\thinnerspace k & 5\thinnerspace515\thinnerspace k & 20:32 & 2.80 &\bf 4.18 \\
\twitter{@Pamela\_Moore13} & 4 & 4\thinnerspace k & 2\thinnerspace k & 54\thinnerspace k & 21:14 & 2.79 &\bf 4.16 \\
\Rowcolor{twitterblue}\twitter{@UserC}\dag & 1\thinnerspace305 & 51\thinnerspace k & 8\thinnerspace k & $<{}$1\thinnerspace k & 22:16 & 6.36 &\bf 0.80\BotStrut \\
\hline
\setarstrut{\footnotesize}%
\multicolumn{8}{>{\footnotesize}r}{*Influence estimate from Eq.~(\ref{eq:individual_impact_causal_estimand}) applied to data\TopStrut} \\
\multicolumn{8}{>{\footnotesize}r}{\dag Anonymized screen names of currently active accounts} \\
\restorearstrut
\end{tabular*}
\end{table}

The precision of the impact estimates on $\bm{\zeta}$ depends
primarily on the variances of the primary effect coefficient $\tau$
and the social effect coefficient $\gamma_1$, because they drive the
total effect of exposures to the sources. The first and second
diagonal terms ($F_{11}$ and $F_{22}$) of the information matrix
quantify the information content on $\tau$ and $\gamma_1$, so
maximizing these two terms leads to smaller sampling variances and
therefore more precise impact estimates. By inspection, these two
terms are maximized when vertices with large expected outcomes
($\lambda$) receive a large amount of peer exposures ($s^{(1)}$). This
happens when sources are on vertices with high out-degrees and
clustered around vertices with large expected outcomes. For parsimony,
this analysis is on the simplest proposed model, but can be easily
generalized to multi-hops and multi-vertex-covariate situations, with
similar resulting intuition.

The results of the causal impact estimation approach on the
\twitter{\#MacronLeaks} narrative network is shown in
Fig.~\ref{fig:influencemacronleaks} and
Table~\ref{tab:macronleaks_screennames}, where the impact of example
participatory accounts are highlighted. Vertex color indicates the
number of times each account tweets \twitter{\#MacronLeaks}; vertex
size is the out-degree (i.e.\ the number of retweets on each
account). Among the accounts with high estimated impact, there exists
independent confirmation on the prominence of the accounts
\twitter{@wikileaks} and \twitter{@JackPosobiec} in pushing the
\twitter{\#MacronLeaks} narrative~\cite{Marantz2017}. A new finding is
the high impact in the social media network discussing the 2017 French
presidential elections of an account known to be tied to foreign
influence operations: \twitter{@Pamela\_Moore13}~\cite{USHPSCI2017}.

These results highlight the ability of causal influence estimation to
infer high impact beyond statistics based upon activity- and
topologically-based statistics, e.g.\ PageRank
centrality~\cite{Langville2005,Peixoto2014}. Compare \twitter{@UserA}
to \twitter{@UserB}, whose statistics are very similar, yet
\twitter{@UserB} is estimated to have much greater impact because it
serves as a bridge into the predominantly French-speaking subgraph
(the cluster seen in the middle of
Fig.~\ref{fig:influencemacronleaks}).  The highly active accounts
\twitter{@UserA} and \twitter{@UserC} tweeted about
\twitter{\#MacronLeaks} many times, but were not estimated to have
high impact.  In contrast, some accounts with only a few
\twitter{\#MacronLeaks} tweets and lower centrality are estimated to
have very high causal impact.  As observed in
Fig.~\ref{fig:influencemacronleaks}, higher out-degree is correlated
with impact, but provides only partial information for influence
estimation. E.g.\ \twitter{@Pamela\_Moore13} has much lower out-degree
count than \twitter{@wikileaks}, \twitter{@JackPosobiec}, or
\twitter{@UserB}, but nevertheless has high impact from her strong
influence on the vertices near her in the retweet network.

Additional analysis is necessary to address important issues and
potential limitations of these specific results and the current
approach. Specifically, this analysis uses a uniquely impactful source
per causal influence estimate; generalizing the source vector would
include the effect of exposure to multiples sources on outcomes.  The
analysis also relies upon a specific potential outcome model. Rigorous
model checking and tuning must be performed.  Because one cannot
depend on knowing the true potential outcome model in real-world
studies, the biasing effects of confounding covariates may not be
removed entirely through modeling.  Additional mitigation should be
done through selecting and weighing observed outcomes to balance the
confounding covariates across different exposure groups to address
biased sampling effects~\cite{Rosenbaum1983}.  Lastly, this analysis
focuses on the population engaged in the narrative. The effect of this
selection bias can be addressed by including the population that is
not engaged, but receptive to influence.  Fortuitously, the narrative
filtering step of Section~\ref{sec:graph-filtering} can provide data
on this type of population.

\section{CONCLUSIONS}
\label{sec:concl}

A novel approach to influence estimation on social media networks is
introduced and applied to the problem of characterizing online
influence operations. The approach uses a network causal inference
framework applied to social media network data arising from graph
sampling and filtering. The causal framework directly infers influence
as a function of exposure.  The Cram\'er-Rao lower bound on the model
parameter covariance is computed, providing insight about the effect of
the social media network on estimation precision.  Twitter data
collected during the 2017 French presidential elections are used to
demonstrate influence estimation performance. A new finding is the
high impact of an account known to be tied to foreign influence
operations in social media discussions of the French
elections. Independent confirmation of the ability to infer high
causal influence is provided from subsequent evidence from the
U.S.~Congress and journalistic reports.

\goodbreak

\let\bibliographysize=\footnotesize 

\bibliographystyle{IEEEbib}
\bibliography{strings,refs}

\endgroup 

\end{document}